\documentclass[11pt,fleqn]{article} 
\usepackage{psfig}
\title{Possible Frictionless Regime for \\ Ultra-High  Temperature Amorphous Matter}  %
\author{Zotin K.-H. Chu$^{1}$}
\date{$^1$Centre for Distribution, P.O. Box 39, Toudiban, Road Xihong, Urumqi 830000,
China} 
\begin{document}           
\maketitle
\doublerulesep=6.5mm        
\baselineskip=6.5mm
\oddsidemargin-1mm         
%
\begin{abstract}
The almost frictionless transport of the very-high temperature
amorphous matter which resembles the  color glass condensate
(possibly having much of their origin in the RHIC heavy ion
collisions) in a confined annular  tube with transversely
corrugations is investigated by using the verified transition-rate
model and boundary perturbation method. We found that for certain
activation volume and energy there exist possible frictionless
states which might be associated with the perfect fluid formation
during the early expansion stage in RHIC Au+Au collisions. We also
address the possible similar scenario in LHC Pb+Pb collisions
considering the possible perfect fluid formation in ultra-high
temperature transport of amorphous matter.
\newline

\end{abstract}
%
\bibliographystyle{plain}
\section{Introduction}
The success of ideal hydrodynamics  [1-7],  in explaining bulk of
the data in Au+Au collisions at RHIC (cf., e.g., [8-9]), has led
to a paradigm that in Au+Au collisions a nearly perfect fluid is
created. There is no {\it a priori} reason why ideal hydrodynamics
works so well at RHIC. The relation between the inverse Reynolds
number ($R^{-1}$) and the attenuation length ($L_s
=(4/3)\eta/(e+p)$) becomes $R^{-1} \approx L_s/\tau$ in the
Bjorken flow case ($\eta$ is the shear viscosity, $e$ is the
energy density, $p$ is the pressure, $\tau (\sim 0.6$ fm/c) is the
expansion rate) [10-11]. $R^{-1}$ is found to be very small from
the blast-wave fitting with the above correction term (cf. [10]).
This suggests that the hadronic fluid in heavy ion collisions is
nearly a perfect one.\newline The perfect fluid can also be
understood from the form of the stress-energy tensor. In the
approximation where the Lagrangian is ${\cal L} = P(X)$, with
$P(X)$ being the kinetic function, the stress-energy tensor has
the form
\begin{equation}
 T_{\mu\nu} \propto -P(X) g_{\mu\nu}+2 P'(X) u_{\mu} u_{\nu},
\end{equation}
where $u_{\mu} =\partial_{\mu} \phi \equiv {\bf u}$, $\phi$ is a
(real) scalar. This has the form of a stress-energy tensor for a
perfect fluid with 4-velocity $u_{\mu}$ [12]. The fact that
$u_{\mu}$ is a gradient means that the flow is irrotational :
$\nabla \times {\bf u} = 0$.
 Note that the defining property of a perfect fluid
 (fluid  with zero viscosity [13]) is that at each point
 there is a frame in which the stress-energy
tensor has the form $T_{\mu\nu} = \mbox{diag}(-\rho, p, p, p)$ and
in ideal hydrodynamics, the mean free path of particles is assumed
to be zero. \newline The interesting (relevant to our borrowed
approach here) issue is the possible existence of the color glass
condensate (CGC) (cf. e.g., [13,14,7,8]). Au+Au collisions are
then collisions of two sheets of colored glass, with the produced
quarks and gluons materializing at a time given by the inverse of
the saturation momentum. The proposed color glass condensate was
thought to be a possible precursor state to the quark-gluon plasma
(QGP). To be specific, at very high energies, multi-particle
production in QCD is generated by small $x$ partons in the nuclear
wavefunctions ($x$ is the fraction of the longitudinal momentum
carried by the parton). These partons have properties best
described as a color glass condensate (CGC) [5]. When two sheets
of CGC collide in a high-energy heavy-ion collision, these partons
are released and create energy densities an order of magnitude
above the energy density required for the crossover from hadronic
to partonic degrees of freedom. This matter, at early times after
a heavy-ion collision, is a coherent classical field, which
expands, decays into nearly on shell partons and may eventually
thermalize to form a quark-gluon plasma. Because it is formed by
melting the frozen CGC degrees of freedom and is the
non-equilibrium matter preceding the QGP, this matter is called
the glasma  [5,13] (a non-equilibrium gluonic state between the
collision moment and equilibrated QGP [6]). \newline As remarked
in [13], the color glass condensate is the matter associated with
the wee partons of a high energy hadronic wavefunction. This
matter has very high energy density. It has properties similar to
Bose-Einstein condensates (there is a characteristic momentum
scale, termed the saturation scale, below which the gluon density
saturates. This effect sets in when $x$ becomes small and the
associated gluon wave length increases to nuclear dimensions. In
such a regime gluons may interact and form a coherent state
reminiscent of a Bose-Einstein condensate) and to spin glasses. It
was motivated based on observations in deep inelastic scattering.
The idea is that as one goes to higher and higher energies, there
are more and more gluons in the hadron wavefunction. They have to
squeeze together, and highly occupy phase space, making a highly
coherent high density system of gluons. This matter controls the
high energy limit of hadronic scattering and provides the initial
conditions for the matter made in such collisions. Thus the almost
perfect fluid phenomena (or frictionless transport) observed
during the early stage (expansion) for Au+Au collisions in RHIC
could be dominated by small $x$ (gluons) (so that the geometrical
scaling occurs) [13,14]. Note that the gluon population at low $x$
is not an incoherent superposition of nucleon structure functions
but is limited with increasing $Au$ by nonlinear gluon-gluon
fusion resulting from the overlap of gluons from several nucleons
in the plane of the nucleus transverse to the collision axis.
\newline
In fact due to the nearly perfect fluid (very low viscosity or
flow resistance being almost zero) characteristics associated to a
possible new state of matter created in Au+Au collisions at RHIC
some researchers argued that this property (the (kinematic) shear
viscosity of this nearly perfect fluid has been determined and
found to be at least a factor of 4 smaller than that of the
superfluid $^4$He [2]) could be called {\it high temperature
superfluidity} because the fluid of quarks in the super-high
temperature range flows better than the superfluid $^4$He in the
extremely low temperature region (cf. [2,4]). With above facts we
can have in mind the possible link between the (color) glassy
condensate (with the role of gluons) and their almost frictionless
transport (of the fluid or liquid). This latter situation is
similar to the proposed superglass reported quite recently
[15,16].
\newline
Meanwhile we know that at low temperatures gases condense into the
liquid (or solid) state. In a liquid transport is no longer
governed by the motion of individual molecules or composite
(condensed) particles. As noticed in [1], Eyring proposed that
momentum transport of condensed composite particles or molecules
(liquid-like) is due to processes that involve the motion of
vacancies [17-19]. These processes can be viewed as thermally
activated transitions in which a molecule or a cluster moves from
one local energy minimum to another. Eyring's approach involves
the Planck constant $h$ and the appearance of $h$ is related to
Eyring's assumption that the collision time of the molecules is
$h/(k_B T)$, the shortest timescale in a liquid ($k_B$ is the
Boltzmann constant, $T$ is the temperature). One important
observation is : Eyring's approach has been successfully applied
to the study of transport of glassy matter at a wide range of
temperatures [17-19]. Possible frictionless transport of glassy
matter under specific environment were also reported recently
[16,20]. \newline
%
Above mentioned analogy or observed facts is the primary
motivation for our present study.
Our concern is  associated with the possible superfluidity (or
perfect fluid) formation after shear-thinning (i.e., the viscosity
diminishes with increasing shear rate.). While superflow (or
perfect fluid flow) in a state of matter possessing a shear
modulus might initially seem untenable, experimental claims for
precisely this phenomenon in solid $^4$He now abound (cf.
[15,16,20].
\newline
In this paper,  to demonstrate the nearly frictionless transport
of CGC and identify the possible transition (or critical)
temperature, we shall adopt the verified transition-rate-state
model [17-19] to study the transport of CGC (presumed to be
amorphous) within a corrugated annular tube (shell-like). The
possible nearly frictionless states due to strong shear-thinning
will be relevant to the perfect fluid formation at the early stage
for Au+Au collidions in RHIC as mentioned above. To obtain the law
of shear-thinning matter for explaining the too rapid annealing at
the earliest time, because the relaxation at the beginning was
steeper than could be explained by the bimolecular law, a
hyperbolic sine law between the shear (strain) rate :
$\dot{\gamma}$ and  shear stress : $\tau$ has been proposed and
the close agreement with experimental data was obtained [17-19].
This model has sound physical foundation from the thermal
activation process [1,17-19] (a kind of quantum tunneling which
relates to the matter rearranging by surmounting a potential
energy barrier was discussed therein). With this model we can
associate the (shear-thinning) fluid with the momentum transfer
between neighboring  clusters on the microscopic scale and reveals
the possible microscopic interaction in the relaxation of flow
with dissipation (the momentum transfer depends on the activation
(shear) volume : $V^*\equiv V_h$ which is associated with the
center distance between atoms and is equal to $k_B T/\tau_0$
($\tau_0$ a constant with the dimension of stress).
\newline To consider the more realistic but complicated confined  boundary
conditions in the interfaces of the annular tube (shell-like),
however, we will use the boundary perturbation technique [16,20]
to handle the presumed wavy-roughness along the interfaces of the
confined annual tube. To obtain the analytical and approximate
solutions, here, the roughness is only introduced in the radial or
transverse direction. The relevant boundary conditions along the
wavy-rough surfaces will be prescribed below. We shall describe
our approach after this section : Introduction with the focus upon
the transition-rate approach and boundary perturbation method. The
approximate expression of the transport is then demonstrated at
the end. Finally, we will illustrate our results into three
figures and give discussions therein.
%
\section{Theoretical Formulations} %
Researchers have been interested in the question of how (complex)
matter responds to an external mechanical load. External loads
cause transport, in Newtonian or various types of non-Newtonian
ways. Amorphous matter, composed of polymers, metals, or ceramics,
can deform under mechanical loads, and the nature of the response
to loads often dictates the choice of matter in various
applications. The nature of all of these responses depends on both
the temperature and loading rate [1].\newline
To the best knowledge of the author, the simplest model that makes
a prediction for the rate and temperature dependence of shear
yielding is the rate-state model of stress-biased thermal
activation [17-19]. Structural rearrangement is associated with a
single energy barrier $E$ that is lowered or raised linearly by an
applied stress $\sigma$ : $R_{\pm}=\nu_0 \exp[-E/(k_B T)] \exp[\pm
\sigma V^*/(k_B T)],$ where $\nu_0$ is an attempt frequency and
$V^*$ is a constant called the 'activation volume'. In amorphous
matter, the transition rates are negligible at zero stress. Thus,
at finite stress one needs to consider only the rate $R_{+}$ of
transitions in the direction aided by stress.\newline The linear
dependence will always correctly describe small changes in the
barrier height, since it is simply the first term in the Taylor
expansion of the barrier height as a function of load. It is thus
appropriate when the barrier height changes only slightly before
the system escapes the local energy minimum. This situation occurs
at higher temperatures; for example, Newtonian transport is
obtained in the rate-state model in the limit where the system
experiences only small changes in the barrier height before
thermally escaping the energy minimum. As the temperature
decreases, larger changes in the barrier height occur before the
system escapes the energy minimum (giving rise to, for example,
non-Newtonian transport). In this regime, the linear dependence is
not necessarily appropriate, and can lead to inaccurate modeling.
To be precise, at low shear rates ($\dot{\gamma} \le
\dot{\gamma}_c$), the system behaves as a power law shear-thinning
material while, at high shear rates, the stress varies affinely
with the shear rate. These two regimes correspond to two stable
branches of stationary states, for which data obtained by imposing
either $\sigma$ or $\dot{\gamma}$ exactly superpose. \newline We
shall consider a steady transport of the  amorphous matter (CGC)
in a wavy-rough annular tube of $r_1$ (mean-averaged inner radius)
with the inner interface being a fixed wavy-rough surface :
$r=r_1+\epsilon \sin(k \theta+\beta)$ and $r_2$ (mean-averaged
outer radius) with the outer interface being a fixed wavy-rough
surface : $r=r_2+\epsilon \sin(k \theta)$, where $\epsilon$ is the
amplitude of the (wavy) roughness, $\beta$ is the phase shift
between two walls, and the roughness wave number : $k=2\pi /L $
($L$ is the wavelength of the surface modulation in
transverse direction). 
\newline Firstly, this amorphous matter (composed of composite condensed
particles, say, quarks and gluons)  can be expressed as [16,18,20]
 $\dot{\gamma}=\dot{\gamma}_0  \sinh(\tau/\tau_0)$,
where $\dot{\gamma}$ is the shear rate, $\tau$ is the shear
stress, $\tau_0 =2 k_B T/V_h$, and $\dot{\gamma}_0 (\equiv C_k k_B
T \exp(-\Delta E/k_B T)/h$) is with the dimension of the shear
rate; here $C_k \equiv 2 V_h/V_m$ is a constant relating rate of
strain to the jump frequency ($V_h=\lambda_2\lambda_3\lambda$,
$V_m=\lambda_2\lambda_3\lambda_1$, $\lambda_2 \lambda_3$ is the
cross-section of the transport unit on which the shear stress
acts, $\lambda$ is the distance  jumped  on each relaxation,
$\lambda_1$ is the perpendicular distance between two neighboring
layers of particles sliding past each other), accounting for the
interchain co-operation required, $h$ is the Planck constant,
$\Delta E$ is the activation energy. In fact, the force balance
gives the shear stress at a radius $r$ as $\tau=-(r \,\delta{\cal
G})/2$ [16,20]. $\delta{\cal G}$ is the net effective
 forcing along the transport (or tube-axis : $z$-axis)
direction (considering $dz$ element).\newline Introducing the
forcing parameter
$\Phi = -(r_2/2\tau_0) \delta{\cal G}$
then we have
 $\dot{\gamma}= \dot{\gamma}_0  \sinh ({\Phi r}/{r_2})$.
As $\dot{\gamma}=- du/dr$ ($u$ is the velocity of the transport in
the longitudinal ($z$-)direction of the annular tube), after
integration, we obtain
\begin{equation}
 u=u_s +\frac{\dot{\gamma}_0 r_2}{\Phi} [\cosh \Phi - \cosh (\frac{\Phi r}{r_2})],
\end{equation}
here, $u_s (\equiv u_{slip})$ is the velocity over the (inner or
outer) surface of the annular (cosmic) string, which is determined
by the boundary condition. We noticed that  a general boundary
condition for transport over a solid surface [16,20] was
\begin{equation}
 \delta u=L_s^0 \dot{\gamma}
 (1-\frac{\dot{\gamma}}{\dot{\gamma}_c})^{-1/2},
\end{equation}
where  $\delta u$ is the velocity jump over the solid surface,
$L_s^0$ is a constant slip length, $\dot{\gamma}_c$ is the
critical shear rate at which the slip length diverges. Note that
the slip (velocity) boundary condition above (related to the slip
length) is closely linked to  the mean free path of the particles
together with a  geometry-dependent factor (in low temperature
regime it is the quantum-mechanical scattering of Bogoliubov
quasiparticles which is responsible for the loss of transverse
momentum transfer to the confined interfaces [21]). The value of
$\dot{\gamma}_c$ is a function of the corrugation of interfacial
energy.  \newline With the slip boundary condition [16,20], we can
derive the velocity fields and transport rates along the
wavy-rough annular (cosmic) string below using the verified
boundary perturbation technique [16,22] and dimensionless
analysis. We firstly select $L_s^0$ to be the characteristic
length scale and set $r'=r/L_s^0$, $R_1=r_1/L_s^0$,
$R_2=r_2/L_s^0$, $\epsilon'=\epsilon/L_s^0$. After this, for
simplicity, we drop all the primes. It means, now, $r$, $R_1$,
$R_2$ and $\epsilon$ become dimensionless ($\Phi$ and
$\dot{\gamma}$ also follow). The wavy interfaces are prescribed as
$r=R_2+\epsilon \sin(k\theta)$ and $r=R_1+\epsilon
\sin(k\theta+\beta)$ and the presumed steady transport is along
the $z$-direction (annulus-axis direction).
\subsection{Boundary Perturbation}
Along the outer interface (the same treatment below could also be
applied to the inner interface), we have
 $\dot{\gamma}=(d u)/(d n)|_{{\mbox{\small on interfaces}}}$.
Here, $n$ means the  normal. Let $u$ be expanded in $\epsilon$ :
 $$u= u_0 +\epsilon u_1 + \epsilon^2 u_2 + \cdots,$$
and on the boundary, we expand $u(r_0+\epsilon dr,
\theta(=\theta_0))$ into
\begin{displaymath}
u(r,\theta) |_{(r_0+\epsilon dr,\,\theta_0)} =u(r_0,\theta)+\epsilon
[dr \,u_r (r_0,\theta)]+ \epsilon^2 [\frac{dr^2}{2}
u_{rr}(r_0,\theta)]+\cdots=
\end{displaymath}
\begin{equation}
  \{u_{slip} +\frac{\dot{\gamma} R_2}{\Phi} [\cosh \Phi - \cosh (\frac{\Phi
 r}{R_2})]\}|_{{\mbox{\small on interfaces}}}, \hspace*{6mm} r_0 \equiv
 R_1, R_2;
\end{equation}
where
\begin{equation}
 u_{slip}|_{{\mbox{\small on interfaces}}}=L_s^0 \{\dot{\gamma}
 [(1-\frac{\dot{\gamma}}{\dot{\gamma}_c})^{-1/2}]\}
 |_{{\mbox{\small on interfaces}}}, 
\end{equation}
Now, on the outer interface (cf., e.g., [16,22])
\begin{displaymath}
 \dot{\gamma}=\frac{du}{dn}=\nabla u \cdot \frac{\nabla (r-R_2-\epsilon
\sin(k\theta))}{| \nabla (r-R_2-\epsilon \sin(k\theta)) |}
=[1+\epsilon^2 \frac{k^2}{r^2}  \cos^2 (k\theta)]^{-\frac{1}{2}}
[u_r |_{(R_2+\epsilon dr,\theta)} -
\end{displaymath}
\begin{displaymath}  
 \hspace*{12mm} \epsilon \frac{k}{r^2}
\cos(k\theta) u_{\theta} |_{(R_2+\epsilon dr,\theta)}
]=u_{0_r}|_{R_2} +\epsilon [u_{1_r}|_{R_2} +u_{0_{rr}}|_{R_2}
\sin(k\theta)-
\end{displaymath}
\begin{displaymath}
  \hspace*{12mm}  \frac{k}{r^2} u_{0_{\theta}}|_{R_2} \cos(k\theta)]+\epsilon^2 [-\frac{1}{2} \frac{k^2}{r^2} \cos^2
(k\theta) u_{0_r}|_{R_2} + u_{2_r}|_{R_2} + u_{1_{rr}}|_{R_2} \sin(k\theta)+ 
\end{displaymath}
\begin{equation}
   \hspace*{12mm} \frac{1}{2} u_{0_{rrr}}|_{R_2} \sin^2 (k\theta) -\frac{k}{r^2}
\cos(k\theta) (u_{1_{\theta}}|_{R_2} + u_{0_{\theta r}}|_{R_2}
\sin(k\theta) )] + O(\epsilon^3 ) .
\end{equation}
Considering $L_s^0 \sim R_1,R_2 \gg \epsilon$ case, we also
presume $\sinh\Phi \ll \dot{\gamma}_c/\dot{\gamma_0}$.
With equations (2) and (6), using the definition of
$\dot{\gamma}$, we can derive the velocity field ($u$) up to the
second order :
\begin{displaymath}
u(r,\theta)=-(R_2 \dot{\gamma}_0/\Phi) \{\cosh
(\Phi r/R_2)-\cosh\Phi\, [1+\epsilon^2 \Phi^2 \sin^2
(k\theta)/(2 R_2^2)]+
\end{displaymath}
\begin{displaymath}
 \hspace*{12mm} \epsilon \Phi \sinh \Phi \,
\sin(k\theta)/R_2\}+u_{slip}|_{r=R_2+\epsilon \sin (k\theta)}.
\end{displaymath}
The key point is to firstly obtain the slip velocity along the
boundaries or surfaces.
After lengthy mathematical manipulations, we obtain %
the velocity fields (up to the second order) and then we can
integrate them with respect to the cross-section to get the transport (volume
flow) rate ($Q$, also up to the second order here) :
\begin{displaymath} 
  Q=\int_0^{\theta_p} \int_{R_1+\epsilon \sin(k\theta+\beta)}^{R_2+\epsilon \sin(k\theta)}
 u(r,\theta) r
 dr d\theta =Q_{0} +\epsilon\,Q_{p_0}+\epsilon^2\,Q_{p_2}.
\end{displaymath}
In fact, the approximate (up to the second order) net transport (volume flow)
rate  reads :
\begin{displaymath}
 Q=\pi \dot{\gamma}_0 \{L_s^0 (R_2^2-R_1^2)  \sinh\Phi \,
 (1-\frac{\sinh\Phi}{\dot{\gamma}_c/\dot{\gamma_0}})^{-1/2}+
 \frac{R_2}{\Phi}[(R_2^2-R_1^2)\cosh\Phi-\frac{2}{\Phi}(R_2^2 \sinh \Phi-
\end{displaymath}
\begin{displaymath}
  R_1 R_2 \sinh(\Phi \frac{R_1}{R_2}))+ \frac{2 R_2^2}{\Phi^2}(\cosh\Phi-\cosh(\Phi \frac{R_1}{R_2}))]\}+
  \epsilon^2 \{\frac{\pi}{2} u_{slip_0} (R_2^2-R_1^2)+
\end{displaymath}
\begin{displaymath}
 L_s^0 \frac{\pi}{4} \dot{\gamma}_0  \sinh \Phi (1+\frac{\sinh\Phi}{\dot{\gamma}_c/\dot{\gamma_0}})
 (-k^2+\Phi^2)[1-(\frac{R_1}{R_2})^2]+\frac{\pi}{2}\dot{\gamma}_0
 [R_1 \sinh (\frac{R_1}{R_2} \Phi)-R_2 \sinh \Phi]-
\end{displaymath}
\begin{displaymath}
 \frac{\pi}{2}  \dot{\gamma}_0 \frac{R_2}{\Phi} [ \cosh\Phi - \cosh (\Phi
  \frac{R_1}{R_2})]+ \frac{\pi}{4}  \dot{\gamma}_0 \Phi \cosh\Phi [R_2  - \frac{R_1^2}{R_2}
  ]+
\end{displaymath}
\begin{displaymath}
  \pi \dot{\gamma}_0 \{[\sinh\Phi+L_s^0 \cosh\Phi
  (1+\frac{\sinh\Phi}{\dot{\gamma}_c/\dot{\gamma_0}})] (R_2-R_1 \cos\beta
  )\}+\frac{\pi}{2}\dot{\gamma}_0 \frac{R_2}{\Phi} \cosh \Phi+
\end{displaymath}
\begin{equation}
  L_s^0\frac{\pi}{4} \Phi^2 \dot{\gamma}_0 \frac{\cosh\Phi}{\dot{\gamma}_c/\dot{\gamma}_0}[1
-(\frac{R_1}{R_2})^2
 ]\} \cosh\Phi.
\end{equation}
Here,
\begin{equation}
 u_{{slip}_0}= L_s^0 \dot{\gamma}_0 [\sinh\Phi(1-\frac{\sinh\Phi}{
 \dot{\gamma}_c/\dot{\gamma}_0})^{-1/2}].
\end{equation}
\section{Results and Discussions}
We firstly check the roughness effect (or combination of curvature
and confinement effects) upon the transport via strongly shearing
because there are no available experimental data and numerical
simulations for the same geometric configuration (annular tube
with wavy corrugations in transverse direction). With a series of
forcings  : $\Phi\equiv - R_2 (\delta {\cal G})/(2\tau_0)$, we can
determine the enhanced shear rates ($d\gamma/dt$) due to gravity
forcings. From equation (6), we have (up to the first order)
\begin{equation}
 \frac{d\gamma}{dt}=\frac{d\gamma_0}{dt} [ \sinh \Phi+\epsilon
 \sin(k\theta) \frac{\Phi}{R_2} \cosh \Phi].
\end{equation}
The parameters
are fixed below (the orientation effect : $\sin(k\theta)$ is fixed
here). $r_2$ (the mean outer radius) is selected as the same as
the slip length $L_s^0$. The amplitude of wavy roughness
can be tuned easily.
The effect of wavy-roughness is significant once the forcing ($\Phi$) is rather
large (the maximum is of the order of magnitude of $\epsilon [\Phi
\tanh(\Phi)/R_2]$). \newline If we select a (fixed) temperature,
 then from the expression of $\tau_0$, we
can obtain the shear stress $\tau$ corresponding to above gravity forcings
($\Phi$) :
\begin{equation}
 \tau =\tau_0 \sinh^{-1} [\sinh(\Phi)+\epsilon
 \sin(k\theta) \frac{\Phi}{R_2} \cosh(\Phi)].
\end{equation}
There is no doubt that the orientation effect ($\theta$) is also
present for the amorphous matter. For illustration below, we only
consider the maximum case : $|\sin(k\theta)|=1$. We shall
demonstrate our transport results below. The wave number of
roughness in transverse direction is fixed to be $10$ (presumed to
be the same for both interfaces of the annular tube) here.
\newline
As the primary interest of present study is related to the
possible frictionless transport  or formation of superfluidity
(presumed to be relevant to the CGC as mentioned in Introduction)
due to strong shearing, we shall present our main results in the
following. Note that, based on the absolute-reaction-rate Eyring
model (of stress-biased thermal activation), structural
rearrangement is associated with a single energy barrier (height)
$\Delta E$ that is lowered or raised linearly by a (shear) yield
stress $\tau$. If the transition rate is proportional to the
plastic (shear) strain rate (with a constant ratio : $C_0$;
$\dot{\gamma} = C_0 R_t$, $R_t$ is the transition rate in the
direction aided by stress), we have
\begin{equation}
 \tau = 2 [\frac{\Delta E}{V_h} + \frac{k_BT}{V_h} \ln(\frac{\dot{\gamma}}{C_0
 \nu_0})]  \hspace*{28mm} \mbox{if}\hspace*{6mm} \frac{V_h
 \tau}{k_B T} \gg 1
\end{equation}
where $\nu_0$ is an attempt frequency or transition rate, $C_0
\nu_0 \sim  \dot{\gamma}_0 \exp(\Delta E/k_B T)$, or
\begin{equation}
 \tau = 2\frac{k_BT}{V_h} \frac{\dot{\gamma}}{C_0
 \nu_0}\exp({\Delta E}/{k_B T})   \hspace*{24mm} \mbox{if}\hspace*{6mm} \frac{V_h
 \tau}{k_B T} \ll 1.
\end{equation}
It is possible that the frictional resistance (or shear stress)
can be almost zero (existence of $\tau \sim 0$) from above
equations (say, equation (11) considering a sudden jump of the
resistance). The nonlinear character only manifests itself when
the magnitude of the applied stress times the activation volume
becomes comparable or greater in magnitude than the thermal
vibrational energy.
\newline Normally, the value of $V_h$ is associated with a typical
volume required for a microscopic shear rearrangement. Thus, the
nonzero transport rate (of the condensed composite (quarks and
gluons) system) as forcing is absent could also be related to a
barrier-overcoming or tunneling for shear-thinning matter along
the wavy-roughness (geometric valley and peak served as potential
surfaces) in annular tubes when the wavy-roughness is present.
Once the geometry-tuned potentials (energy) overcome this barrier,
then the tunneling (spontaneous transport) inside wavy-rough
annular tubes occurs.
Now, we start to examine the temperature effect. We fix the
forcing $\Phi$ to be $1$ as its effect is of the order $O(1)$ for
the shear rate. As the gravity forcing ($\delta {\cal G}$) might
depend on the temperature ($|\delta {\cal G}|=2\tau_0 \Phi/R_2$,
$\tau_0\equiv \tau_0 (T)$, $V^*(\equiv V_h)$ is presumed to be
temperature independent here for simplicity).
Note that, according to [7], $V^*=3 V \delta \gamma/2$
for certain matter during an activation event [7], where $V$ is
the deformation volume, $\delta\gamma$ is the increment of shear
strain.
\newline
As the primary interest of present study is related to the
possible phase transition [14-16] or formation of superfluidity
(presumed to be relevant to the formation of dark matter mentioned
in Introduction) due to strong shearing, we shall present our main
results in the following. We performed intensive calculations or
manipulations of related physical and geometric parameters,
considering a hot big-bang universe [14] and examine what happens
as it expands and cools through the transition temperature $T_c$.
The selected temperature range and the activation anergy follows
this reasoning.  Note that in unified models of weak and
electromagnetic interactions $T_c$ is of the order of the square
root of the Fermi coupling constant [14], $G_F^{1/2}$, i.e. a few
hundred GeV. Thus the transition occurs when the universe is aged
between $10^{-10}$ and $10^{-12}$ seconds and far above nuclear
densities [23]. One possible high-temperature superfluidity
formation (or nearly frictionless transport) regime [2,4] is
demonstrated in Fig. 1.
The activation energy ($\Delta E$) is $6 \times 10^{-10}$ Joule.
In fact, all the results shown in this figure depend on
$\dot{\gamma}_0$ and are thus very sensitive to $\Delta E$ (and
$V_h$).
Here $C_k=2$ and the sudden jump of the shear stress (directly
linked to the friction) occurring around $T\sim 10^{12}$
$^{\circ}$K  could be the transition temperature for the selected
$\Delta E$ and $C_k$. There is a sudden friction drop around  two
orders of magnitude below $T\sim 2 \times 10^{12}$ $^{\circ}$K
($V_h \sim 4 \times 10^{-11}$ m$^3$) and it is almost frictionless
below $T\sim 10^{12}$ $^{\circ}$K. The temperature regime we
identify here is close to the critical (at the critical end point)
as well as crossover (to the quark-gluon plasma (QGP) at zero
baryon chemical potential,) temperature mentioned in [2] (cf.
[1]).
\newline The possible reasoning for this formation can be
illustrated in Fig. 2. It could be due to the strong shearing
driven by larger forcings along a confined tube. The
shear-thinning (the viscosity diminishes with increasing shear
rate) reduces the viscosity significantly. One possible outcome
for almost vanishing viscosity is the nearly frictionless
transport.\newline Based on  the knowledge gained at RHIC, it
should be possible to predict experimental results at the Large
Hadron Collider (LHC), which will collide lead ions at much higher
energy densities [24]. At this point it's important to predict
something about the LHC experiments [13,24] even the LHC will
bring unanticipated discovery. Fig. 3 illustrates the possible
(almost) perfect fluid formation or frictionless transport for
ultra-high temperature transport of amorphous matter (possible
color-glass-condensate [24]) considering Pb+Pb collisions at LHC.
The activation energy is $10^{-9}$ Joule. We can observe there is
a sudden friction (shear stress) drop around two orders of
magnitude below $T\sim 7 \times 10^{17}$ $^{\circ}$K ($V_h \sim
2.5 \times 10^{-6}$ m$^3$) and it is almost frictionless below
$T\sim 3 \times 10^{17}$ $^{\circ}$K.
\section{Conclusions}
To conclude in brief, we have  obtained critical parameters for
the possible perfect fluid formation or almost frictionless
transport of ultra-high temperature amorphous matter (possible CGC
or string-like composite (condensed) particles associated with
voids) relevant to Au+Au collisions at RHIC as well as Pb+Pb
collisions at LHC. These critical parameters depend strongly upon
the temperature, activation energy and activation volume. We shall
investigate other relevant  issues [1,13,25,26] in the future.

\newpage

\psfig{file=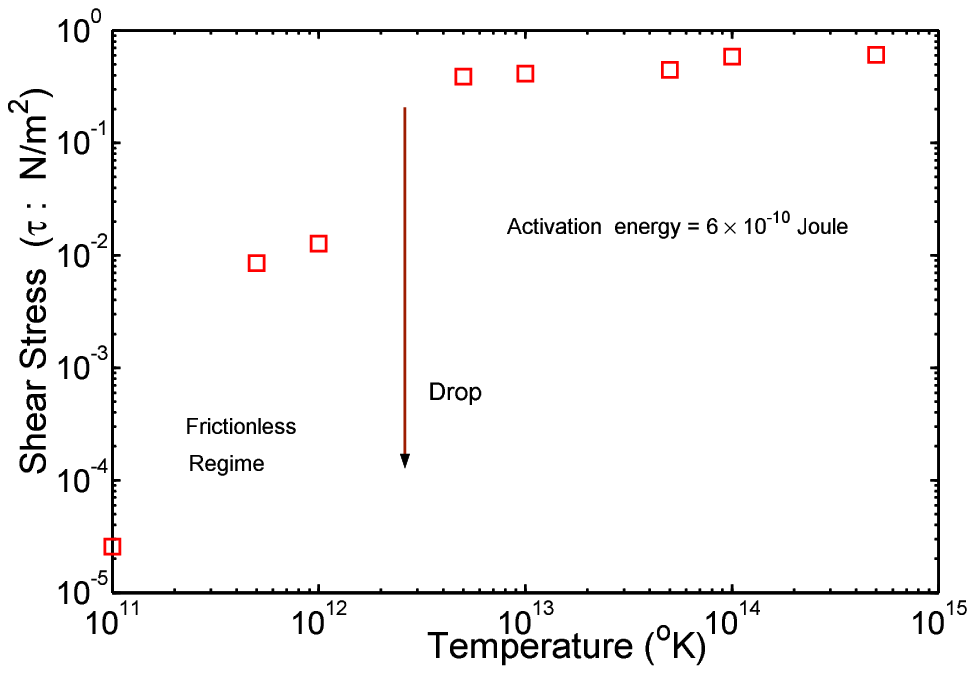,bbllx=-1.5cm,bblly=18cm,bburx=12cm,bbury=26cm,rheight=8cm,rwidth=12cm,clip=}

\begin{figure} [h]
\hspace*{7mm} Figure 1. Comparison of calculated (shear) stresses
using an activation energy
 $6 \times 10^{-10}$ J. \newline \hspace*{8mm} There is a sharp
decrease of shear stress around T $\sim 2 \times 10^{12}$ K.
 Below $10^{12}$ K, \newline \hspace*{8mm} the transport of amorphous matter is nearly
frictionless (cf. [1,2] for the critical  \newline \hspace*{8mm}
temperatures related to Au+Au collisions at RHIC).
\end{figure}

\newpage

\psfig{file=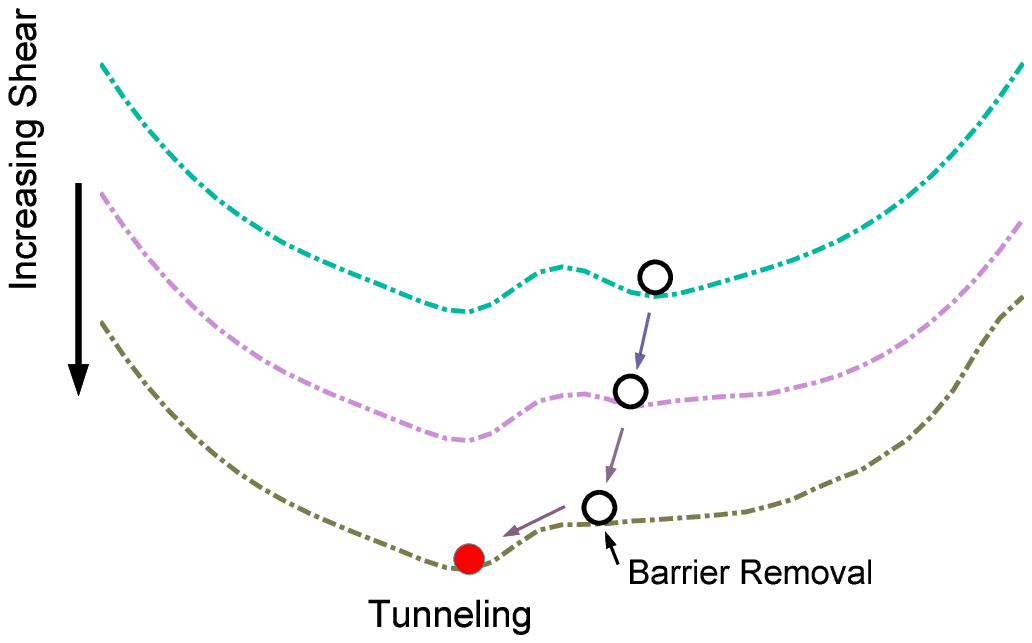,bbllx=-1.0cm,bblly=18cm,bburx=12cm,bbury=26cm,rheight=8cm,rwidth=10cm,clip=}

\begin{figure} [h]
\hspace*{8mm} Figure 2. Increasing shear causes a local energy
minimum to flatten until it disappears
\newline  \hspace*{9mm} (energy barrier removal or quantum-like
tunneling). The structural contribution \newline  \hspace*{9mm}
to the shear stress is referred to shear-thinning.
\end{figure}

\newpage

\psfig{file=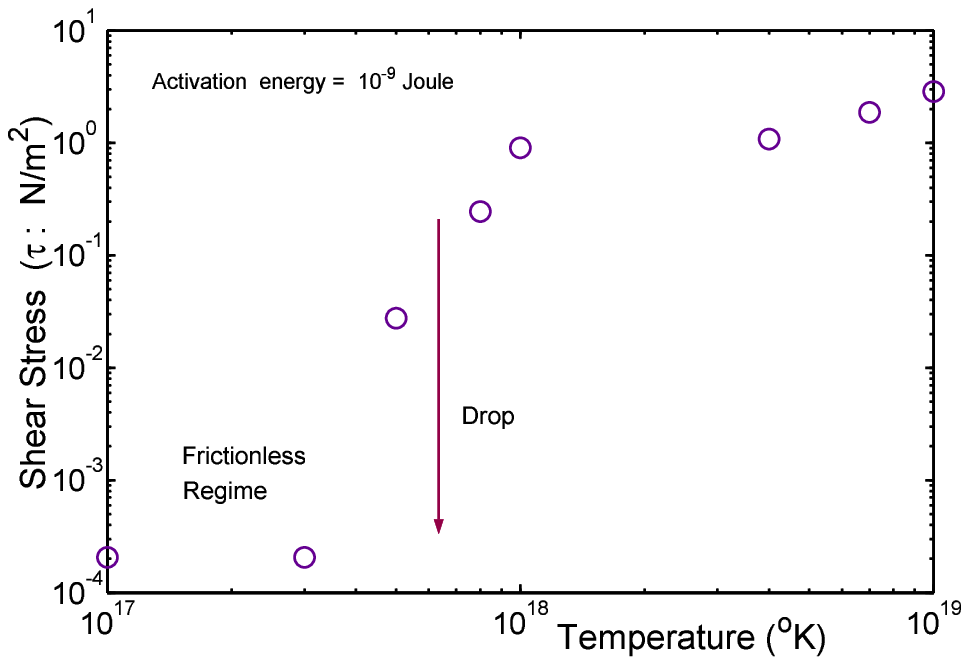,bbllx=-1.5cm,bblly=18cm,bburx=12cm,bbury=26cm,rheight=8cm,rwidth=12cm,clip=}

\begin{figure} [h]
\hspace*{7mm} Figure 3. Comparison of calculated (shear) stresses
using an activation energy
 $10^{-9}$ J. \newline \hspace*{8mm} There is a sharp
decrease of shear stress around T $\sim 7 \times 10^{17}$ K.
 Below $3 \times 10^{17}$ K, \newline \hspace*{8mm} the transport of amorphous matter is nearly
frictionless (cf. [24] for the possible critical  \newline
\hspace*{8mm} temperatures in Pb+Pb collisions at LHC).
\end{figure}
\end{document}